\documentclass[preprint2]{aastex}

\usepackage{graphicx}

\newcommand{\sm}[1]{\mbox{{\scriptsize #1}}}

\newcommand{\bef}{\begin{figure}}
\newcommand{\eef}{\end{figure}}

\def\eps@scaling{0.95}

\def\showone#1{
  \centering
  \leavevmode
  \epsfxsize=\eps@scaling\linewidth
  \epsfbox{#1.eps}
}

\def\showtwover#1#2{
  \centering
  \leavevmode
  \epsfxsize=\eps@scaling\linewidth
  \epsfbox{#1.eps} \hfil
  \epsfxsize=\eps@scaling\linewidth
  \epsfbox{#2.eps}
}

\def\showthreeover#1#2#3{
  \centering
  \leavevmode
  \epsfxsize=\eps@scaling\linewidth
  \epsfbox{#1.eps} \hfil
  \epsfxsize=\eps@scaling\linewidth
  \epsfbox{#2.eps} \hfil
  \epsfxsize=\eps@scaling\linewidth
  \epsfbox{#3.eps}
}

\def\showfourover#1#2#3#4{
  \centering
  \leavevmode
  \epsfxsize=\eps@scaling\linewidth
  \epsfbox{#1.eps} \hfil
  \epsfxsize=\eps@scaling\linewidth
  \epsfbox{#2.eps} \hfil
  \epsfxsize=\eps@scaling\linewidth
  \epsfbox{#3.eps} \hfil
  \epsfxsize=\eps@scaling\linewidth
  \epsfbox{#4.eps}
}

\def\epstwo@scaling{0.46}

\def\showtwo#1#2{
  \centering
  \leavevmode
  \epsfxsize=\epstwo@scaling\linewidth
  \epsfbox{#1.eps} 
  \epsfxsize=\epstwo@scaling\linewidth
  \epsfbox{#2.eps}
}

\def\epsthree@scaling{0.28}
\def\showthree#1#2#3{
  \centering
  \leavevmode
  \epsfysize=\epsthree@scaling\textwidth 
  \epsfbox{#1.eps} 
  \epsfysize=\epsthree@scaling\textwidth 
  \epsfbox{#2.eps}
  \epsfysize=\epsthree@scaling\textwidth 
  \epsfbox{#3.eps}
}

\def\epstwo@scaling{0.44}

\def\showfour#1#2#3#4{
  \centering
  \leavevmode
  \epsfxsize=\epstwo@scaling\linewidth
  \epsfbox{#1.eps} \hfil
  \epsfxsize=\epstwo@scaling\linewidth
  \epsfbox{#2.eps} \hfil
  \epsfxsize=\epstwo@scaling\linewidth
  \epsfbox{#3.eps} \hfil
  \epsfxsize=\epstwo@scaling\linewidth
  \epsfbox{#4.eps}
}

\def\showsix#1#2#3#4#5#6{
  \centering
  \leavevmode
  \epsfxsize=\epstwo@scaling\linewidth
  \epsfbox{#1.eps} \hfil
  \epsfxsize=\epstwo@scaling\linewidth
  \epsfbox{#2.eps} \hfil
  \epsfxsize=\epstwo@scaling\linewidth
  \epsfbox{#3.eps} \hfil
  \epsfxsize=\epstwo@scaling\linewidth
  \epsfbox{#4.eps} \hfil
  \epsfxsize=\epstwo@scaling\linewidth
  \epsfbox{#5.eps} \hfil
  \epsfxsize=\epstwo@scaling\linewidth
  \epsfbox{#6.eps}
}

\newcommand{\befone}{
  \begin{figure*}
  \centering
  \begin{minipage}{\textwidth}
  }
\newcommand{\eefone}{\end{minipage}\end{figure*}}

\newcommand{\HI}{$\mathrm{H}$\ }

\newcommand{\HzI}{$\mathrm{H}_2$\ }

\newcommand{\HM}{$\mathrm{H}^-$\ }

\newcommand{\HeI}{$\mathrm{He}$\ }
\newcommand{\HeII}{$\mathrm{He}^+$\ }

\newcommand{\HDI}{$\mathrm{HD}$\ }

\newcommand{\HId}{$\mathrm{H}$}
\newcommand{\HIId}{$\mathrm{H}^+$}
\newcommand{\HzIId}{$\mathrm{H}_2^+$}
\newcommand{\HzId}{$\mathrm{H}_2$}

\newcommand{\HMd}{$\mathrm{H}^-$}

\newcommand{\HeHIId}{$\mathrm{HeH}^+$}

\newcommand{\DId}{$\mathrm{D}$}
\newcommand{\DIId}{$\mathrm{D}^+$}
\newcommand{\HDIId}{$\mathrm{HD}^+$}
\newcommand{\HDId}{$\mathrm{HD}$}
\newcommand{\DMd}{$\mathrm{D}^-$}

\newcommand{\fHeI}{\mathrm{He}}

\newcommand{\fe}{\mathrm{e}}
\newcommand{\fpp}{\mathrm{p}}

\shorttitle{Magnetic fields and the first stars}
\shortauthors{Schleicher et al.}

\begin{document}


\title{The influence of magnetic fields on the thermodynamics of primordial star formation}


\author{Dominik R. G. Schleicher\altaffilmark{1}, Daniele Galli\altaffilmark{2}, Simon C. O. Glover\altaffilmark{1}, Robi Banerjee\altaffilmark{1}, \\Francesco Palla \altaffilmark{2}, Raffaella Schneider\altaffilmark{2}, Ralf~S.~Klessen\altaffilmark{1}}
\email{dschleic@ita.uni-heidelberg.de, galli@arcetri.astro.it, sglover@ita.uni-heidelberg.de, banerjee@ita.uni-heidelberg.de, palla@arcetri.astro.it, raffa@arcetri.astro.it, rklessen@ita.uni-heidelberg.de}

\altaffiltext{1}{Zentrum f\"ur Astronomie der Universit\"at Heidelberg, Institut f\"ur Theoretische Astrophysik, Albert-Ueberle-Str. 2, D-69120 Heidelberg, Germany}
\altaffiltext{2}{INAF - Osservatorio Astrofisico di Arcetri, Largo Enrico Fermi 5, I-50125 Firenze, Italia}

\begin{abstract}
We explore the effects of magnetic energy dissipation on the formation of the first stars. For this purpose, we follow the evolution of primordial chemistry in the presence of magnetic fields in the post-recombination universe until the formation of the first virialized halos. From the point of virialization, we follow the protostellar collapse up to densities of $\sim10^{12}$~cm$^{-3}$ in a one-zone model. In the intergalactic medium (IGM), comoving field strengths of $\gtrsim0.1$~nG lead to Jeans masses of $10^8\ M_\odot$ or more and thus delay gravitational collapse in the first halos until they are sufficiently massive. During protostellar collapse, we find that the temperature minimum at densities of $\sim10^{3}$~cm$^{-3}$ does not change significantly, such that the characteristic mass scale for fragmentation is not affected. However, we find a significant temperature increase at higher densities for comoving field strengths of $\gtrsim0.1$~nG. This may delay gravitational collapse, in particular at densities of $\sim10^{9}$~cm$^{-3}$, where the proton abundance drops rapidly and the main contribution to the ambipolar diffusion resistivity is due to collisions with Li$^+$. We further explore how the thermal evolution depends on the scaling relation of magnetic field strength with density. While the effects are already significant for our fiducial model with $B\propto\rho^{0.5-0.57}$, the temperature may increase even further for steeper relations and lead to the complete dissociation of \HzI at densities of $\sim10^{11}$~cm$^{-3}$ for a scaling with $B\propto\rho^{0.6}$. The correct modeling of this relation is therefore very important, as the increase in temperature enhances the subsequent accretion rate onto the protostar. Our model confirms that initial weak magnetic fields may be amplified considerably during gravitational collapse and become dynamically relevant. For instance, a comoving field strength above $10^{-5}$~nG will be amplified above the critical value for the onset of jets which can magnetize the IGM.

\end{abstract}

\section{Introduction}
In the absence of magnetic fields, protostellar collapse during first star formation is well understood and has been followed in detail by numerical simulations \citep{Abel02, Bromm04}. The main physical processes, including chemistry, cooling, the extent of fragmentation and protostellar feedback have been discussed in detail by \citet{Ciardi05} and \citet{Glover05}. Recent numerical simulations confirmed this scenario with unprecedented accuracy \citep{Yoshida08}. The first stars are expected to be very massive (with characteristic masses of $\sim100\ M_\odot$), consistent with constraints from the observed reionization optical depth \citep{Wyithe03, SchleicherBanerjee08}.

However, it may be important to consider the effects of putative magnetic fields. Recent observations by \citet{Bernet08} showed that strong magnetic fields ($B\sim3\mu$G) were present in normal galaxies at $z\sim3$, where relatively little time was available for a dynamo to operate. In our galaxy, the field strength is $3-4\ \mathrm{\mu}$G, and it is coherent over kpc scales, with alternating directions in the arm and interarm regions \citep[e.g.][]{Kronberg94, Han08}. Such alternations are expected for magnetic fields of primordial origin, but more difficult to explain in most dynamo models \citep{Grasso01}. Moreover, it is not clear whether large-scale dynamos are efficient, as the small-scale magnetic fields are produced at a faster rate and lead to saturation before a significant large-scale field builds up \citep{Kulsrud97}. \citet{Zweibel06} confirms problems with explaining the observed magnetic field strength from dynamo theory in our galaxy, while models based on stellar magnetic fields have difficulties with explaining the large-scale coherence. On the other hand, magnetic fields observed in some other spiral galaxies appear to be in agreement with the predictions of dynamo theory \citep{Beck09}. The main arguments against primordial magnetic fields have been considered by \citet{Kulsrud08}, who find that they are too uncertain to rule out this possibility. {Strong magnetic fields have also been detected in the polarized disk in M~31 \citep{Berkhuijsen03}.}

At redshifts higher than $z>3$, it is only possible to derive upper limits on the magnetic field strength. Observations of small-scale cosmic microwave background (CMB) anisotropy yield an upper limit of $4.7$~nG to the comoving field strength \citep{Yamazaki06}. Additional constraints can be derived from the measurement of $\sigma_8$, which describes the root-mean-square of the density fluctuations at $8h^{-1}$~comoving~Mpc \citep{Yamazaki08}. The inferred field strength is similar, but depends on the assumed shape of the power spectrum. Primordial nucleosynthesis constrains the comoving field strength to be less than $\sim1$~$\mathrm{\mu}$G \citep{Grasso96}, while reionization yields upper limits of $0.7-3$~nG, depending on assumptions on the stellar population that is responsible for reionizing the universe \citep{SchleicherBanerjee08}. As shown in previous works, these upper limits are relatively weak, but may imply a significant impact on the thermal evolution in the postrecombination universe and structure formation \citep{Wasserman78, Kim96, Sethi05, Tashiro06a, Tashiro06b, Sethi08, SchleicherBanerjee08, SchleicherBanerjee09a}.  {Numerical simulations based on an SPH-MHD version of Gadget indicate that a comoving field strength of $2\times10^{-3}$~nG might be sufficient to reproduce the observed magnetic fields in galaxy clusters \citep{Dolag02, Dolag05}. However, larger initial field strengths also appear to be consistent with the observed data, which shows a large degree of scatter \citep[see Fig.~10 in][]{Dolag02}. Because these results may also change when including processes like AGN feedback, cooling and heat conduction, more work is required before being able to draw definite conclusions about the upper limit to the field strength. 
}

For the early universe, a number of viable scenarios exist that can produce primordial fields. Mechanisms to generate such fields in the epoch of inflation have been proposed by \citet{Turner88}, and more recent works confirm that strong magnetic fields of up to $1$~nG~(comoving) may be produced in this epoch \citep{Bertolami99, Bamba08, Campanelli08a, Campanelli08b}. These models require that conformal invariance is broken explicitly or implicitly \citep{Parker68}. Magnetic fields may also form during the electroweak phase transition \citep{Baym96}, especially if it is a phase-transition of first order, as required in the electroweak baryogenesis scenario \citep{Riotto99}. Similarly, the QCD phase transition may give rise to strong primordial fields \citep{Quashnock89, Cheng94}. Particularly strong fields ($\sim1$~nG, comoving) can be generated if hydrodynamical instabilities were present at this epoch, as expected for a large range of model parameters \citep{Sigl97}. The further evolution of magnetic fields generated at the QCD or electroweak phase transition has been calculated by \citet{Banerjee04b}, who find that the coherence length may increase further if the magnetic helicity is non-zero. 

There is thus a reasonable possibility that magnetic fields were present in the universe after recombination and during the formation of the first stars. Previous works have started to explore the effects of magnetic fields, both those of primordial origin, as well as those created during structure formation or in the protostellar disk. \citet{Maki04, Maki07} investigated the coupling  of the magnetic field to the primordial gas during collapse, finding that the field is frozen to the gas and may drive the magnetorotational instability (MRI) in the accretion disk. \citet{Tan04} proposed that gravitational instabilities in the accretion disk may give rise to turbulence and drive a dynamo that increases the field strength until equipartition is reached. \citet{Silk06} discussed whether the MRI can drive a dynamo in the protostellar disk, suggesting that magnetic feedback may give rise to a distribution of stellar masses closer to the present-day initial mass function (IMF). The effects of such feedback have been explored by \citet{Machida06}, who find that jets may blow away up to $10\%$ of the accreting matter. On the other hand, \citet{Xu08} calculated the generation of magnetic fields by the Biermann battery mechanism \citep{Biermann50} during the formation of the first stars, and find that such fields are not strong enough to affect the dynamics of first star formation. Simulations of present-day star formation indicate that magnetic fields suppress fragmentation of cluster-forming molecular cores \citep{Hennebelle08, Price08} and extract a considerable amount of angular momentum from the protostellar disk \citep[e.g.,][]{Banerjee07a} helping to form more massive stars. {Massive stars may play an important role in creating and spreading magnetic fields on timescales of a few million years \citep{Kogan73a, Kogan73b, Kogan77}. As recently shown by \citet{Hanasz09}, the cosmic ray-driven dynamo proposed by \citet{Parker92} could work on timescales faster than the rotation timescale of a galaxy, and therefore be important even for the earliest galaxies \citep{Bouwens06, Iye06}.}

The effects of magnetic fields on the first stars are therefore still subject to significant uncertainties. In this work, we point out that magnetic fields may not only directly affect the dynamics by launching jets or driving the MRI in the accretion disk, as previously suggested, but also indirectly through changes in the chemistry of the gas and additional heat input from magnetic energy dissipation. We focus on comoving field strengths of up to $\sim1$~nG, as stronger fields are difficult to generate in the early universe and current constraints would not allow for their existence. We calculate the evolution of primordial chemistry in the presence of magnetic fields until the formation of virialized halos in \S~\ref{IGM}. In \S~\ref{collapse}, we follow the thermal evolution of the gas during collapse in a one-zone model and show the impact of the magnetic field on the thermal and magnetic Jeans mass.  The implications of our results are discussed in \S~\ref{discussion}.

\section{From the large-scale IGM to virialized halos}\label{IGM}
Here, we explore the effects of magnetic fields on the chemical evolution in the post-recombination universe until the formation of the first virialized halos at $z\sim20$. This calculation provides the initial conditions used for our models of protostellar collapse that is described in \S~\ref{collapse}.

\subsection{Model description}
We calculate the thermal evolution of the IGM in the postrecombination universe with a modified version of the RECFAST code\footnote{http://www.astro.ubc.ca/people/scott/recfast.html} \citep{Seager99}, a simplified but accurate version of a detailed code that follows hundreds of energy levels for \HId, \HeI and \HeII and self-consistently calculates the background radiation field \citep{Seager00}. The recombination calculation was recently updated by \citet{Wong08}, and the physics of magnetic fields in the IGM and of reionization have been added by \citet{SchleicherBanerjee08}. Here we extend this model to follow the thermal evolution until virialization and to self-consistently follow the chemical evolution until that point. At high redshift $z>40$, the universe is close to homogeneous, and the evolution of the temperature $T$ is given as
\begin{eqnarray}
\frac{dT}{dz}&=&\frac{8\sigma_T a_R T_{\sm{rad}}^4}{3H(z)(1+z)m_e c}\frac{x_e\,(T-T_{\sm{rad}})}{1+f_{\fHeI}+x_e}\nonumber\\
&+&\frac{2T}{1+z}-\frac{2(L_{\sm{AD}}-L_{\sm{cool}})}{3nk_B H(z)(1+z)},\label{temp}
\end{eqnarray}
where $L_{\sm{AD}}$ is the heating term due to ambipolar diffusion (AD), $L_{\sm{cool}}$ is the cooling term including Lyman $\alpha$, bremsstrahlung and recombination cooling based on the cooling functions of \cite{Anninos97}. In Eq.~(\ref{temp}), $\sigma_T$ is the Thomson scattering cross section, $a_R$ the Stefan-Boltzmann radiation constant, $m_e$ the electron mass, $c$ the speed of light, $k_B$ Boltzmann's constant, $n$ the total number density, $x_e=n_\fe/n_H$ the electron fraction per hydrogen atom, $T_{\sm{rad}}$ the CMB temperature, $H(z)$ is the Hubble factor and $f_{\fHeI}$ is the number ratio of \HeI and \HI nuclei. The latter can be obtained as $f_{\fHeI}=Y_p/4(1-Y_p)$ from the mass fraction $Y_p$ of \HeI with respect to the total baryonic mass. The AD heating rate is given as \citep{Pinto08a}
\begin{equation}
 L_{\sm{AD}}=\frac{\eta_{\sm{AD}}}{4\pi}\left|\left(\nabla\times\vec{B}\right)\times\vec{B}/B\right|^2,\label{ambiheat}
\end{equation}
where $\eta_{\sm{AD}}$ is given as
\begin{equation}
 \eta_{\sm{AD}}^{-1}=\sum_n \eta_{\sm{AD},n}^{-1}.\label{etaAD}
\end{equation}
In this expression, the sum includes all neutral species $n$, and $\eta_{\sm{AD},n}$ denotes the AD resistivity of the neutral species $n$. We calculate these resistivities based on the multifluid approach described in Appendix B of \citet{Pinto08a}. This approach is a more general version of the formalism used by \citet{SchleicherBanerjee08}, and is particularly convenient to calculate the AD heating rate in the high-density regime in \S~\ref{collapse}. 

{Using this approach, we have calculated both the ambipolar diffusion heating rate as well as the Ohmic heating rate from the Hall parameters $\beta_{sn}$, defined as}
\begin{equation}
\beta_{sn}=\left( \frac{q_s B}{m_s c}\right)\frac{m_s+m_n}{\rho_n \langle \sigma v \rangle_{sn}}.
\end{equation}
{In this context, the index $s$ denotes ionized species and $n$ neutral species. $q_s$ is therefore the charge of the ionized particle, $m_s$ its mass, $m_n$ the mass of the neutral particle, $\rho_n$ the density of the neutral species, and  $\langle \sigma v \rangle_{sn}$ the zero drift velocity momentum transfer coefficients for these species (the zero drift approximation holds in the absence of shocks).}

{
 From these Hall parameters, it is possible to calculate the conductivity parallel to the electric field, $\sigma_{||}$, as well as the Pedersen and Hall conductivities, $\sigma_P$ and $\sigma_H$, respectively. In general, these are calculated as sums over different pairs of ions and neutrals. In this paragraph, we focus only on the dominant contribution, which typically comes from from collisions between the most abundant ion and neutral species. Then, the resistivities scale as $\sigma_{||}\propto\beta_{sn}$, $\sigma_P\propto\beta_{sn}/(1+\beta_{sn}^2)$ and $\sigma_H\propto1/(1+\beta_{sn}^2)$.  We further checked that $\beta_{sn}>>1$ in the different regimes of our calculation. Therefore, $\sigma_{||}>>\sigma_P>>\sigma_H$. For this reason, the ambipolar diffusion resistivity scales as $\eta_{AD}\propto(\sigma_P/(\sigma_P^2+\sigma_H^2)-1/\sigma_{||})\propto1/\sigma_P$, while the Ohmic resistivity scales as $\eta_O\propto1/\sigma_{||}$. We therefore find that $\eta_{AD}>>\eta_O$. We have checked this in more detail by evaluating the ambipolar diffusion and Ohmic heating rates during the calculation. As ambipolar diffusion always appeared to be the dominant magnetic energy dissipation mechanism in the cases considered here, we will focus on this contribution in the subsequent discussion. Further details regarding the multifluid approach can be found in \citet{Pinto08a}.
}

In the IGM, the dominant contributions to the total resistivity are the resistivities of atomic hydrogen and helium due to collisions with protons. For their calculation, we adopt the momentum transfer coefficients of \citet{Pinto08b}. 

As the power spectrum of the magnetic field is unknown and Eq.~(\ref{ambiheat}) cannot be solved exactly, we adopt a simple and intuitive approach to estimate the differential operator for a given average magnetic field $B$ with coherence length $L_B$. The heating rate can then be evaluated as 
\begin{equation}
L_{\sm{AD}}\sim\frac{\eta_{\sm{AD}}}{4\pi}\frac{B^2}{L_B^2}.\label{ambiheatapprox}
\end{equation}
In the IGM, one can show that $\eta_{\sm{AD}}\propto B^2$, so that we recover the same dependence on $B$ and $L_B$ as \citet{SchleicherBanerjee08}. The coherence length $L_B$ is in principle a free parameter that depends on the generation mechanism of the magnetic field and its subsequent evolution. It is constrained through the fact that tangled magnetic fields are strongly damped by radiative viscosity in the pre-recombination universe on scales smaller than the Alfv{\'e}n damping scale $k_{\sm{max}}^{-1}$  given by \citep{Jedamzik98, Subramanian98, Seshadri01}
\begin{eqnarray}
k_{\sm{max}}&\sim&234\ \mathrm{Mpc}^{-1}\left(\frac{B_0}{1\ \mathrm{nG}} \right)^{-1}\left(\frac{\Omega_m}{0.3}\right)^{1/4}\nonumber\\
&\times&\left(\frac{\Omega_b h^2}{0.02}\right)^{1/2}\left(\frac{h}{0.7} \right)^{1/4},\label{minlength}
\end{eqnarray}
where $B_0=B/(1+z)^2$ denotes the comoving magnetic field, $\Omega_m$ and $\Omega_b$ are the cosmological density parameters for the total and baryonic mass, and $h$ is the Hubble constant in units of $100$~km~s$^{-1}$~Mpc$^{-1}$. In fact, we expect that fluctuations of the magnetic field may be present on larger scales as well. We thus estimate the heating rate by adopting $L_B=k_{\sm{max}}^{-1}(1+z)^{-1}$, which yields the dominant contribution. \citet{Sethi05} also considered contributions from decaying MHD turbulence. However, these contributions were found to be negligible compared to the AD heating terms, and highly uncertain, since only mild and sub-Alfv{\'e}nic turbulence can be expected at these early times. Even in the first star-forming halos, numerical hydrodynamics simulations show that the gas flow is laminar and only a small amount of turbulence is present \citep{Abel02, Bromm04}. We therefore neglect this contribution here. 

The additional heat input provided by AD affects the evolution of the ionized fraction $x_\fpp$ of hydrogen, which is given as
\begin{eqnarray}
\frac{dx_\fpp}{dz}&=&\frac{[x_e x_p n_H \alpha_H -\beta_H (1-x_p)e^{-h_p\nu_{H, 2s}/kT}] }{H(z)(1+z)[1+K_H(\Lambda_H+\beta_H)n_H(1-x_p)]}\nonumber\\
&\times&[1+K_H\Lambda_H n_H(1-x_p)]-\frac{k_{\sm{ion}}n_H x_p}{H(z)(1+z)}.\label{ion}
\end{eqnarray}
Here, $n_H$ is the number density of hydrogen atoms and ions, $h_p$ Planck's constant, $k_{\sm{ion}}$ is the collisional ionization rate coefficient 
 \citep{Abel97}. For the further details of notation, see \citet{SchleicherBanerjee08, Seager99}. The parametrized case B recombination coefficient for atomic hydrogen $\alpha_H$ is given by
\begin{equation}
\alpha_H=F\times10^{-13}\frac{at^b}{1+ct^d}\ \mathrm{cm}^3\ \mathrm{s}^{-1}
\end{equation} 
with $a=4.309$, $b=-0.6166$, $c=0.6703$, $d=0.5300$ and $t=T/10^4\ K$, which is a fit given by \citet{Pequignot91}. This coefficient takes into account that direct recombination into the ground state does not lead to a net increase in the number of neutral hydrogen atoms, since the photon emitted in the recombination process can ionize other hydrogen atoms in the neighbourhood.  The fudge factor $F=1.14$ serves to speed up recombination and is determined from comparison with the multilevel code. 

The chemical evolution of the primordial gas is solved with a system of rate equations for \HMd, \HzIId, \HzId, \HeHIId, \DId, \DIId, \DMd, \HDIId and \HDId, which is largely based on the reaction rates presented by \citet{Schleicher08}. For the mutual neutralization rate of \HM and \HIId, we use the more recent result of \citet{Stenrup09}. We note that the collisional dissociation rates of \HzI in this compilation are somewhat larger than those of \citet{Stancil98}. For this reason, the final \HzI abundance in our calculation is smaller by up to an order of magnitude than in the calculation of \citet{Sethi08}, which was based on the rates of \citet{Stancil98}. The evolution of the magnetic field strength can be determined from the magnetic field energy $E_B=B^2/8\pi$ and is followed as
\begin{equation}
\frac{dE_B}{dt}=\frac{4}{3}\frac{\partial\rho}{\partial t}\frac{E_B}{\rho}-L_{\sm{AD}}.\label{bfield}
\end{equation}
The first term describes the evolution of the magnetic field {in a homogeneous universe in the absence of specific magnetic energy generation or dissipation mechanisms.  If this term dominates, then the magnetic field strength evolves as $B\propto\rho^{2/3}$ due to the expansion of space, and the magnetic energy scales with redshift as $(1+z)^4$, the same scaling as for the radiation energy density.  } The second term accounts for corrections due to energy dissipation via AD. Such energy dissipation is reflected in the evolution of the magnetic Jeans mass, which in general is defined as
\begin{equation}
 M_J^B=\frac{\Phi}{2\pi\sqrt{G}},\label{mgJeans}
\end{equation}
with the magnetic flux $\Phi=\pi r^2 B$, where $G$ is Newton's constant, and an appropriate length scale $r$. For the IGM, a characteristic Jeans length can be derived from the Alfv\'en velocity and the dynamical timescale, yielding  \citep{Subramanian98, Sethi05}
\begin{equation}
M_J^B\sim10^{10}M_\odot\left(\frac{B_0}{3\ \mathrm{nG}} \right)^3.
\end{equation}
This equation is formally independent of density, as it is expressed here in terms of comoving quantities. The constant comoving IGM density is thus absorbed in the overall normalization. The additional heat input affects the evolution of the thermal Jeans mass as well, which we evaluate as
\begin{equation}
M_J=\left( \frac{4\pi\rho}{3} \right)^{-1/2}\left(\frac{5k_B T}{2\mu G m_P}\right)^{3/2}.\label{thJeans}
\end{equation}
Here, $k_B$ denotes Boltzmann's constant, $\mu$ the mean molecular weight and $m_p$ the proton mass.

To describe virialization in the first minihalos, we employ the spherical collapse model for pressureless dark matter until an overdensity of $\sim200$ is reached. For this purpose, we first calculate cosmic time in the universe as
\begin{equation}
 t_{\sm{cosmic}}(z)=\frac{2}{3H_0 \sqrt{\Omega_\Lambda}}\log\left(\frac{1+\cos\theta}{\sin\theta} \right),\label{timeun}
\end{equation}
where $H_0$ is the Hubble constant at $z=0$, $\Omega_\Lambda$ the cosmological density parameter for vacuum energy, and $\theta$ is defined via
\begin{equation}
 \tan\theta=\sqrt{\frac{1-\Omega_\Lambda}{\Omega_\Lambda}}\left(1+z\right)^{3/2}.
\end{equation}
On the other hand, in the spherical collapse model, we have the following system of equations \citep{Peebles93}:
\begin{eqnarray}
R&=&A_1(1-\cos\eta),\\
 t&=&A_2(\eta-\sin\eta),\label{timecoll}\\
A_1^3&=&GMA_2^2.\label{consistent}
\end{eqnarray}
Here, $R$ is the radius of the collapsing cloud, $t$ is the time parameter, $A_1$ and $A_2$ are parameters that normalize the length and time scales in this problem, $M$ the total mass, which we choose as $M=10^6\ M_\odot$. (The following results hardly depend on this choice). With $A_2=t_{20}/2\pi$, where $t_{20}$ is the age of the universe at $z=20$, we ensure that the cloud collapses at $z=20$. Eq.~(\ref{consistent}) then fixes the remaining parameter $A_1$. Equating Eq.~(\ref{timeun}) and Eq.~(\ref{timecoll}) allows one to derive the parameter $\eta$ and to calculate the overdensity $\rho/\rho_b$ in the protocloud. The latter is given as
\begin{equation}
 \frac{\rho}{\rho_b}=\frac{9}{2}\frac{(\eta-\sin\eta)^2}{(1-\cos\eta)^3}\left[4-\frac{9}{2}\frac{\sin\eta(\eta-\sin\eta)}{(1-\cos\eta)^2}\right]^{-1}.
\end{equation}
We assume that the formation of the protocloud may affect the coherence length $L_B$ of the magnetic field if it is frozen into the gas. We therefore adopt $L_B=\mathrm{min}(R,k_{\sm{max}}^{-1}(1+z)^{-1})$ at this evolutionary stage.

\subsection{Results}

\begin{figure}[t]
\includegraphics[scale=0.5]{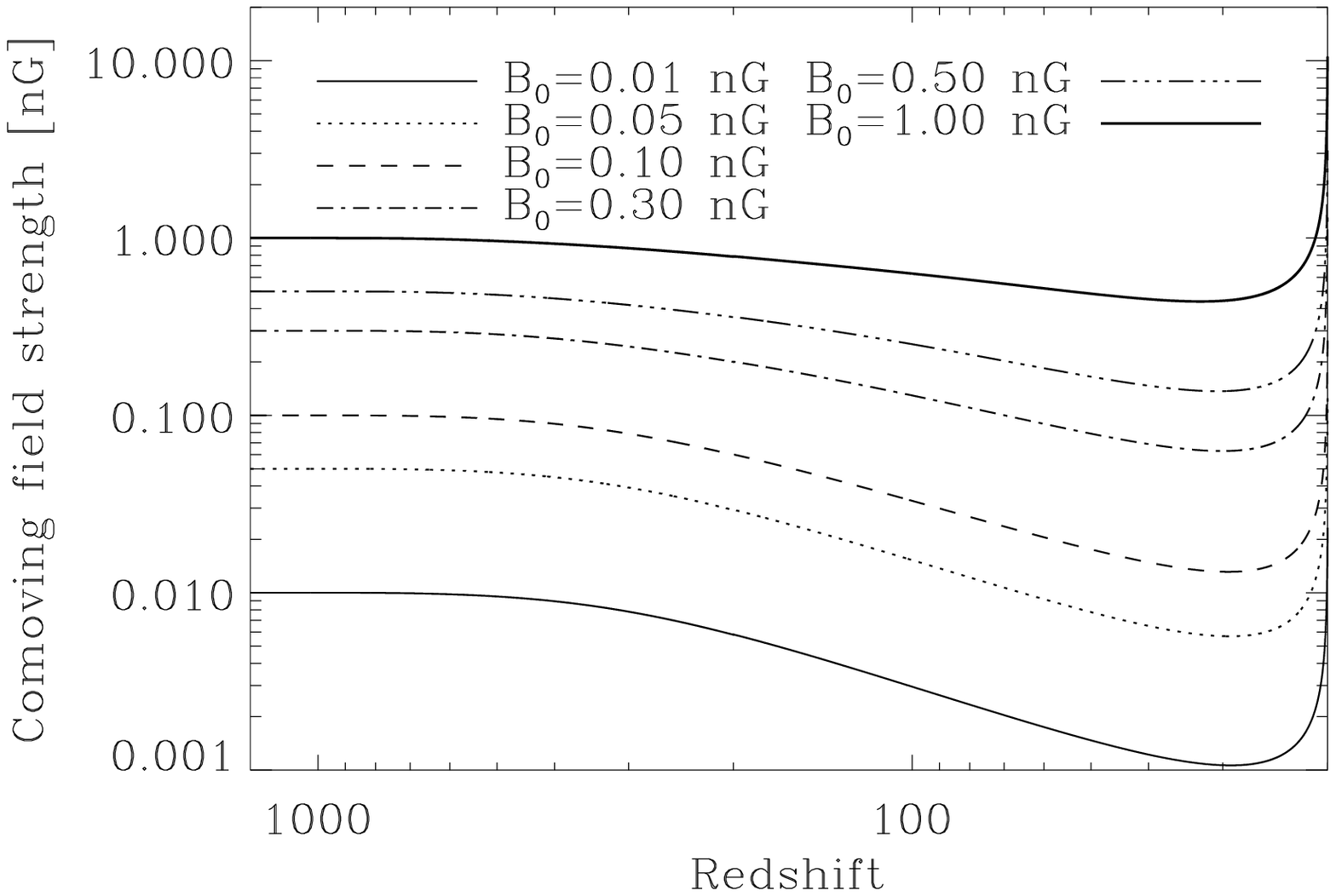}
\caption{The evolution of the comoving magnetic field strength due to AD as a function of redshift for different initial comoving field strengths, from the homogeneous medium at $z=1300$ to virialization at $z=20$.}
\label{fig:bfieldIGM}
\eef

 \begin{figure}[t]
 \includegraphics[scale=0.5]{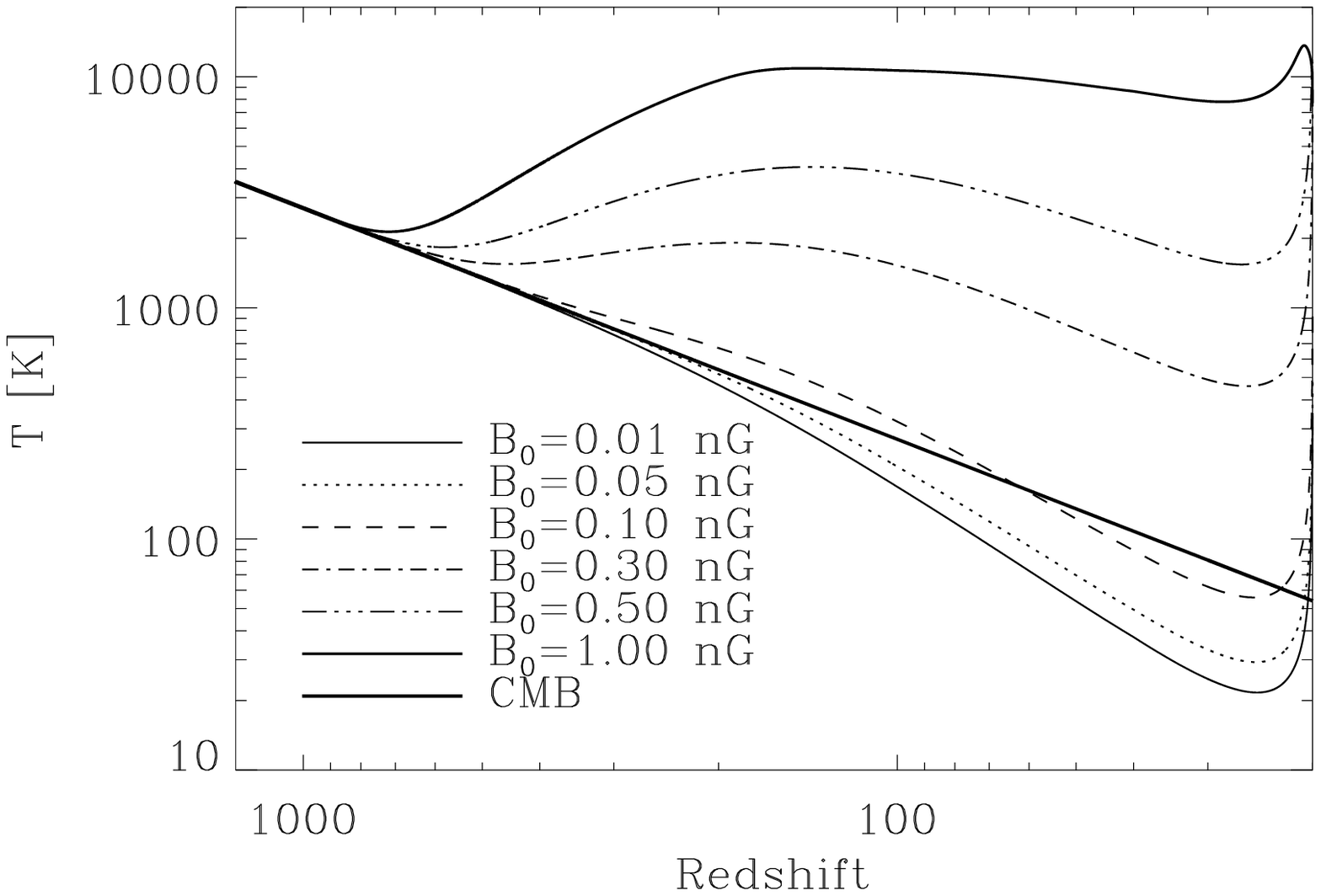}
 \caption{The gas temperature evolution in the IGM as a function of redshift for different comoving field strengths, from the homogeneous medium at $z=1300$ to virialization at $z=20$. For the case with $B_0=0.01$~nG, we find no difference {in the thermal evolution} compared to the zero-field case. }
 \label{fig:tempIGM}
 \eef

\begin{figure}[t]
\includegraphics[scale=0.5]{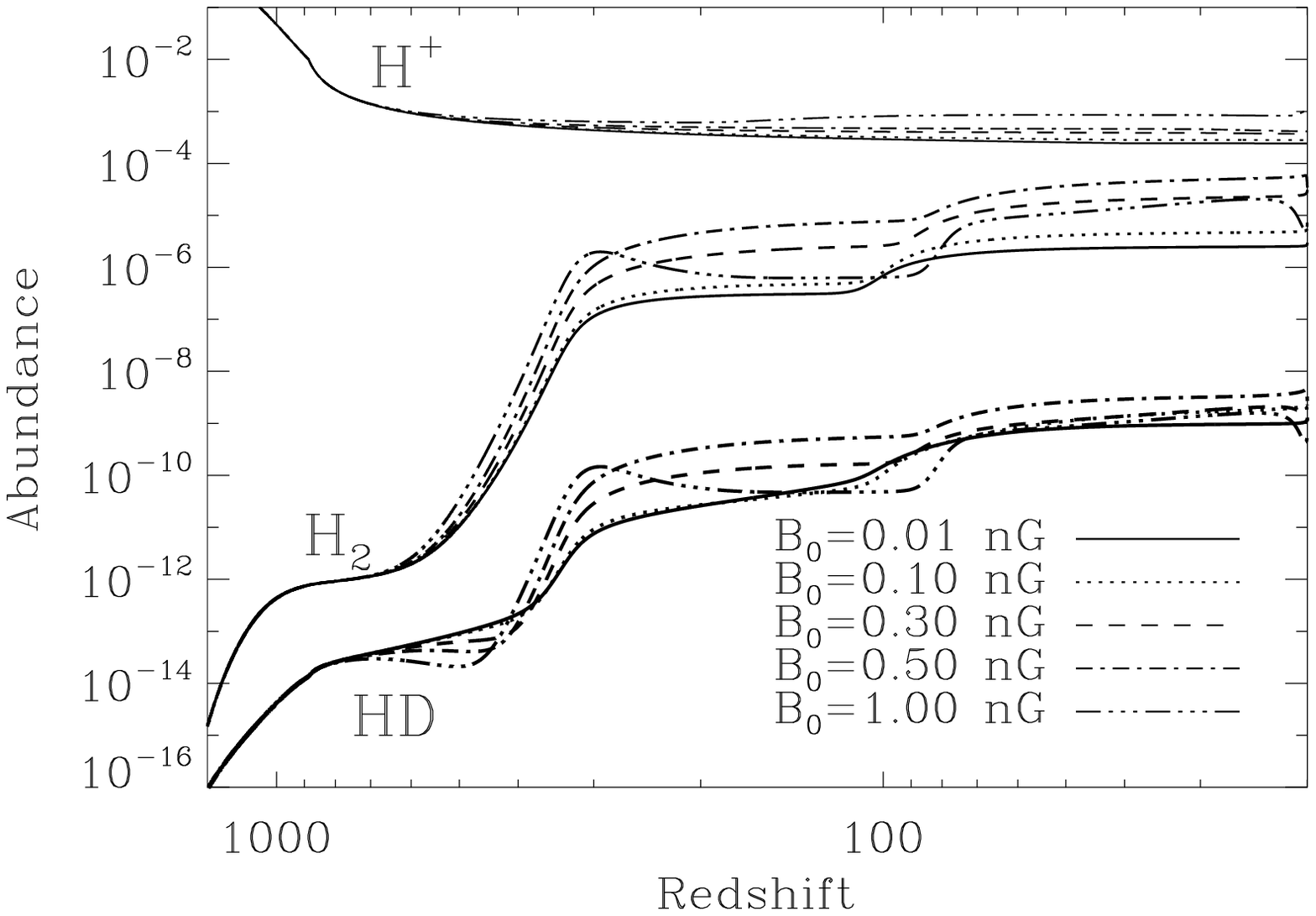}
\caption{The evolution of ionization degree, \HzI and \HDI abundances as a function of redshift for different comoving field strengths, from the homogeneous medium at $z=1300$ to virialization at $z=20$. For the case with $B_0=0.01$~nG, we find no difference {in the chemical evolution} compared to the zero-field case.}
\label{fig:ionIGM}
\eef

With the model described above, we calculate the evolution of the temperature and chemistry of the IGM from $z=1300$ until virialization in the first halos at $z=20$ for different initial comoving field strengths. We note that in our model, the comoving field strength is a function of time, as the magnetic field is dissipated through AD. In the figures given here, $B_0$ therefore labels the initial comoving field strength used to initialize the calculation.

As shown in Fig.~(\ref{fig:bfieldIGM}), ambipolar diffusion is particularly important for fields with initial comoving field strengths of $0.2$~nG or less. For stronger fields, the dissipation of only a small fraction of their energy increases the temperature and the ionization fraction of the IGM to such an extent that AD becomes less effective. The evolution of temperature and ionization degree is given in Figs.~(\ref{fig:tempIGM}) and (\ref{fig:ionIGM}). While for comoving field strengths up to $\sim0.1$~nG, the additional heat from ambipolar diffusion is rather modest and the gas in the IGM still cools below the CMB temperature due to adiabatic expansion, it can increase significantly for stronger fields and reaches $\sim10^4$~K for a comoving field of $1$~nG. At that temperature, Lyman $\alpha$ cooling is efficient, and collisional ionization increases the ionization degree, which makes AD less effective and prevents a further increase in temperature. Fig.~(\ref{fig:ionIGM}) also shows the evolution of the \HzI and \HDI abundances. As a result of the enhanced ionization fraction, they are increased for non-zero field strengths at the point of virialization. For relatively weak fields, the abundances are increased at all redshifts $z<700$ with respect to the zero-field case. For field strengths above $\sim0.7$~nG, the heating is so strong that collisional dissociation of \HM and \HzI is very efficient and the abundances of \HzI and \HDI may even drop below the zero-field value. After this point, however, AD is shut down by the increased ionization fraction and the gas temperature decreases somewhat. The molecules thus reform and the abundances again become larger than in the zero-field case.

\begin{figure}[t]
\includegraphics[scale=0.5]{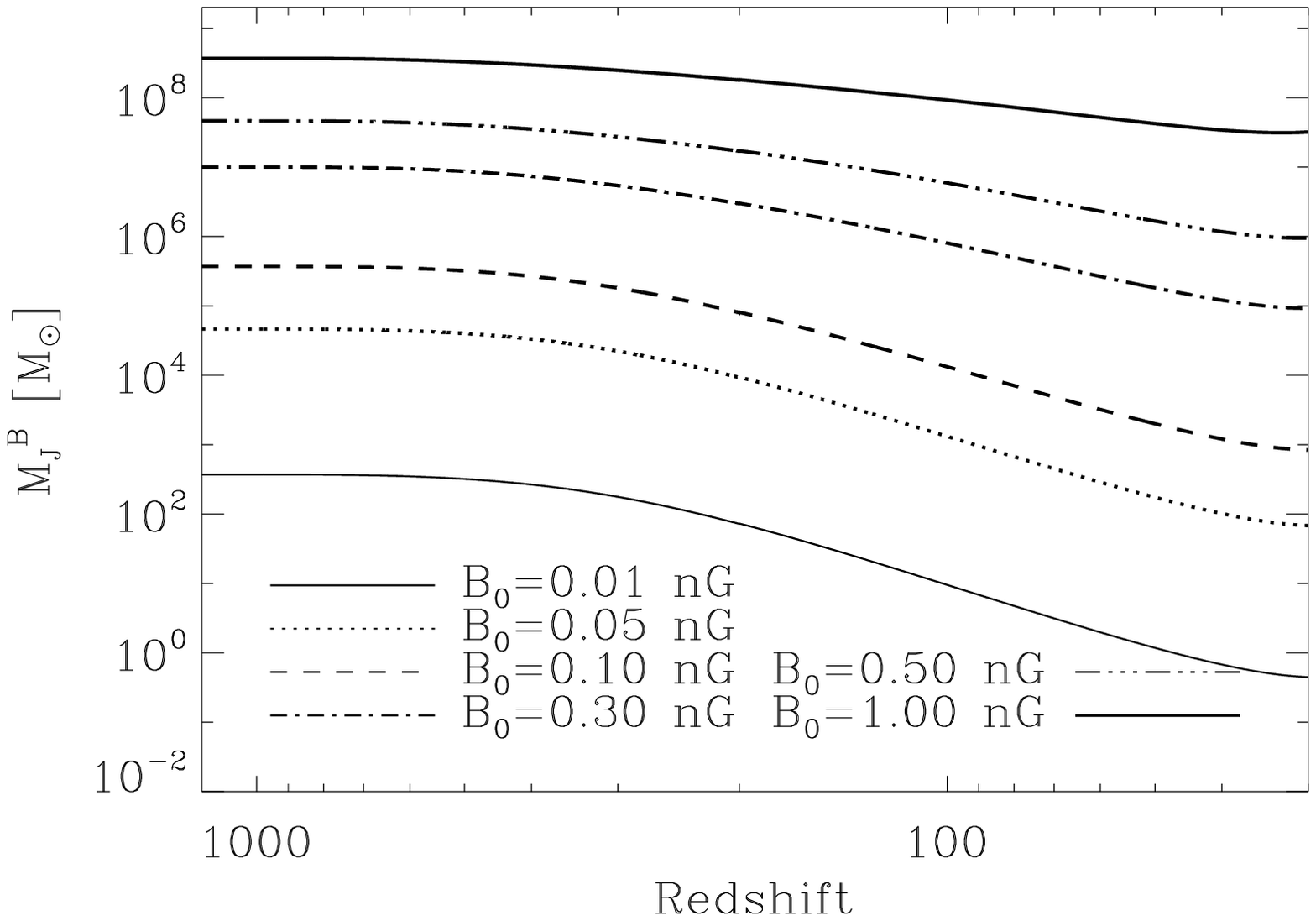}
\caption{The evolution of the magnetic Jeans mass as a function of redshift for different comoving field strengths, from the homogeneous medium at $z=1300$ to virialization at $z=30$.}
\label{fig:mjeansIGM}
\eef

\begin{figure}[t]
\includegraphics[scale=0.5]{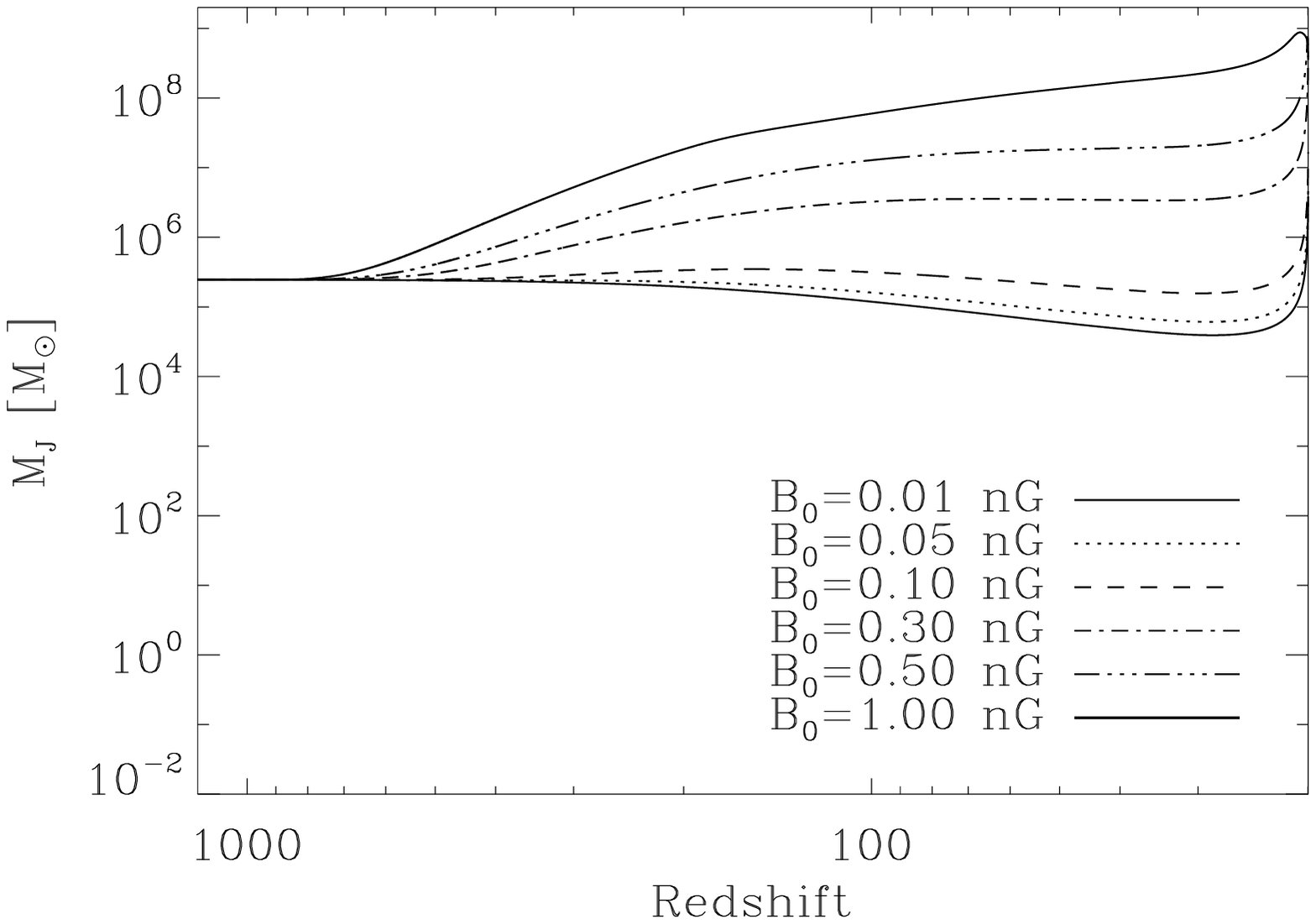}
\caption{The evolution of the thermal Jeans mass as a function of redshift for different comoving field strengths, from the homogeneous medium at $z=1300$ to virialization at $z=20$. For the case with $B_0=0.01$~nG, we find no difference {in the thermal Jeans mass} compared to the zero-field case.}
\label{fig:thjeansIGM}
\eef

The evolution of the comoving field strength is also reflected in the magnetic Jeans mass, as can be seen in Fig.~(\ref{fig:mjeansIGM}). Depending on the field strength, $M_J^B$ may become as large as $\sim10^8\ M_\odot$, thus suppressing star formation in the first minihalos. Due to the steep scaling with the comoving field to the third power, it decreases quickly if the magnetic field is weaker. Also the thermal Jeans mass is affected in the presence of magnetic fields due to the increase of the gas temperature. The impact on $M_J$ is plotted in Fig.~(\ref{fig:thjeansIGM}). The increase is significant for comoving field strengths above $0.1$~nG. Up to $\sim1$~nG, the thermal Jeans mass dominates over the magnetic one. For stronger fields, the gas temperature cannot increase much further, as AD is shut down, while the magnetic Jeans mass would still scale with $B_0^3$ and would therefore dominate. 

In summary, our results show that the presence of magnetic fields may affect the chemical initial conditions for star formation in the first halos at $z\sim20$. They further indicate that the magnetic and/or the thermal Jeans mass may be considerably increased for comoving fields of at least $0.1$~nG, implying that small halos need to accrete more mass until they can eventually collapse.

\section{The protostellar collapse phase}\label{collapse}
In this section, we use the results from the previous section, in particular the chemical initial conditions and the dissipation of the magnetic energy, as input parameters to calculate the chemistry during the collapse of the first protostars. This calculation therefore assumes that the corresponding halos are massive enough, such that the gas is gravitationally unstable and can form stars. If the halo is less massive initially, the gas would linger at low densities until enough material is accreted to go into collapse. Such accretion could be due to minor or major mergers as well as cooling flow activity.

\subsection{Model assumptions}

To follow the evolution of the primordial gas during the protostellar collapse phase, we employ and extend the one-zone model developed by \citet{Glover09}. This model assumes that the collapse occurs on the free-fall timescale, unaffected by changes in the thermal energy of the gas. It incorporates the most extensive treatment of primordial gas chemistry currently available, involving $392$ reactions amongst $30$ atomic and molecular species, and hence allows us to follow the fractional ionization of the gas accurately even for values as small as $x_{\sm{e}}\sim10^{-12}$.  The chemical rate equations are evolved simultaneously with the thermal energy equation, and the coupled set of equations are solved implicitly using the DVODE integrator \citep{Brown89}. The effects of chemical heating due to three-body \HzI formation and chemical cooling due to \HzI and \HDI dissociation are included, and play an important role in regulating the temperature of the gas at densities $n>10^8$~cm$^{-3}$. Particularly important for this application is the fact that it correctly models the evolution of the ionization degree and the transition at densities of $\sim10^8$~cm$^{-3}$ where Li$^+$ becomes the main charge carrier. Further details of the model can be found in \citet{Glover09}.

The additional heat input due to magnetic energy dissipation may however have an important impact on the thermal evolution of the gas and delay gravitational collapse. It may thus lead to a backreaction on the heating rate for gravitational contraction, as for a given change in volume, the energy input from $pdV$ work is fixed, while the amount of energy that is radiated away increases if collapse is delayed. At the same time, the amount of energy dissipated by AD may be increased. To model the complex interplay of the heating and cooling rates involved, we therefore need to evaluate how the collapse timescale is affected by the thermodynamics of the gas. For this purpose, we adopt the approach of \citet{Omukai05} based on the Larson-Penston type self-similar solution \citep{Larson69, Penston69} as generalized by \citet{Yahil83}. In this solution, the actual collapse timescale $t_{\sm{coll}}$ is related to the free-fall timescale $t_{\sm{ff}}$ as
\begin{equation}
 t_{\sm{coll}}=\frac{1}{\sqrt{1-f}}t_{\sm{ff}},
\end{equation}
where $f$ describes the ratio between the pressure gradient force and the gravitational force at the center. It can be calculated from an effective equation of state parameter $\gamma=d\log p/d\log\rho$ as
\begin{eqnarray}
 f&=&0, \qquad  \gamma<0.83, \\
&=&0.6+2.5(\gamma-1)-6.0(\gamma-1)^2, \quad 0.83<\gamma<1,\nonumber\\
&=& 1.0+0.2(\gamma-4/3)-2.9(\gamma-4/3)^2, \quad \gamma>1.\nonumber
\end{eqnarray}
At $\gamma=4/3$, $f$ reaches unity and the collapse timescale diverges, indicating that the self-similar solution is no longer valid under these circumstances. In reality, a hydrostatic core would form and thereafter contract during its further evolution. As in \citet{Omukai05}, we mimic this effect by adopting an upper limit $f=0.95$.

During protostellar collapse, magnetic fields are typically found to scale as a power-law with density $\rho$. Assuming ideal MHD with flux freezing and spherical collapse, one expects a scaling with $\rho^{2/3}$ in the case of weak fields. Deviations from spherical symmetry such as expected for dynamically important fields give rise to shallower scalings, e.g. $B \propto \rho^{0.6}$ \citep{Banerjee06}, $B \propto \rho^{1/2}$ \citep{Spitzer98, Hennebelle08a, Hennebelle08}. This is because collapse preferentially occurs along the field lines in the latter case, and is slowed down in the perpendicular direction. 

The difference in these scaling behaviors may seem negligible at first sight, and is indeed hard to distinguish in simulations that often cover only a few orders of magnitude in density. For our calculations, however, it is an important issue, as they cover about $12$ orders of magnitude in density. As the simulations of \citet{Machida06} cover a similar range of densities and magnetic field strengths as our calculations, we use them to derive an appropriate scaling relation. Of course, this relation is only approximate, as their simulation does not include a consistent treatment of non-equilibrium chemistry and the additional heat input from AD, which may have a backreaction on the dynamics. In their results, the slope $\alpha$ of the scaling relation $B\propto\rho^\alpha$ depends on the ratio of the thermal Jeans mass to the magnetic Jeans mass, which we calculate from Eqs.~(\ref{mgJeans})~and~(\ref{thJeans}).  We find the {empirical scaling law}
\begin{equation}
 \alpha=0.57\left(\frac{M_J}{M_J^B} \right)^{0.0116}.\label{fit}
\end{equation}
{This relation matches our expectations described above. For negligible magnetic pressure and low magnetic Jeans masses, the magnetic field strength increases more rapidly with density, while for strong magnetic pressure, gas collapse occurs preferentially along the magnetic field lines, and the scaling relation flattens. Even though the relation depends only weakly on this ratio, } one should note that $M_J/M_J^B$ may vary by several orders of magnitude depending on the initial field strength. {As an additional caveat, we point out that this relation may to some extent depend on the initial conditions and change depending on the individual properties of the first minihalos. For this reason, we will not only explore the consequences of this particular relation, but also consider the effects from softer or steeper relations below. }

The calculation of the magnetic Jeans mass as given in Eq.~(\ref{mgJeans}) requires an assumption regarding the length scale $r$ of the dense region. Numerical hydrodynamics simulations show that regions of a given density have length scales comparable to the thermal Jeans length \citep{Abel02, Bromm04}. We expect similar effects from the magnetic Jeans mass and therefore take the maximum of the thermal and magnetic Jeans length. As shown in the previous section, the thermal Jeans mass dominates over the magnetic Jeans mass for comoving field strengths up to $\sim1$~nG. We therefore initialize this length scale with the thermal Jeans length at the beginning of collapse, and follow its further evolution by taking the maximum of the thermal and the magnetic Jeans length.

At every timestep, the magnetic field strength is updated as
\begin{equation}
B_{\sm{new}}=B_{\sm{old}}\left(\frac{\rho_{\sm{new}}}{\rho_{\sm{old}}} \right)^\alpha.\label{increaseB}
\end{equation}
The coefficient $\alpha$ is calculated at every timestep such that a change in the ratio of the thermal and magnetic Jeans mass, either due to chemistry, magnetic energy dissipation or magnetic field amplification, may lead to a backreaction on the increase of the field strength with density. However, Eq.~(\ref{increaseB}) does not incorporate the effect of magnetic energy dissipation on the field strength. As for the IGM, we can calculate the AD heating rate from Eq.~(\ref{ambiheatapprox}). As we are here considering a much larger range of densities, additional processes need to be taken into account to calculate the AD resistivity correctly. In particular, at a density of $\sim10^8$~cm$^{-3}$, the three-body \HzI formation rates start to increase the \HzI abundance significantly, such that the gas is fully molecular at densities of $\sim10^{11}$~cm$^{-3}$. As a further complication, the proton abundance drops considerably at densities of $\sim10^8$~cm$^{-3}$, such that Li$^+$ becomes the main charge carrier \citep{Glover09}. To model the total AD resistivity correctly, we therefore need to calculate the resistivities of \HId, \HeI and \HzI and take into account collisions with protons and Li$^+$. For collisions with protons, we adopt the momentum transfer coefficients of \citet{Pinto08a}, while we use the polarization approximation for collisions with Li$^+$ \citep[see][]{Pinto08b}. 

With these considerations, we can calculate the AD heating rate and its impact on the thermodynamics. We evaluate its impact on the magnetic field strength as
\begin{equation}
 B_{\sm{cor}}=\sqrt{8\pi}\sqrt{B_{\sm{uncor}}^2/8\pi-\delta t L_{\sm{AD}}},
\end{equation}
where $\delta t$ denotes the timestep, and $B_{\sm{cor}}$ and $B_{\sm{uncor}}$ denote the magnetic field corrected and not corrected for the effects of AD, respectively. The equation follows from the dependence of the magnetic energy density on the field strength and the impact of AD on the magnetic energy density. Finally, we checked whether Ohmic diffusion \citep[see][]{Pinto08a} can be relevant under primordial conditions, finding that AD is always the dominant source of energy dissipation.

\subsection{Results}\label{ResultsCollapse}

\begin{figure}[t]
\includegraphics[scale=0.5]{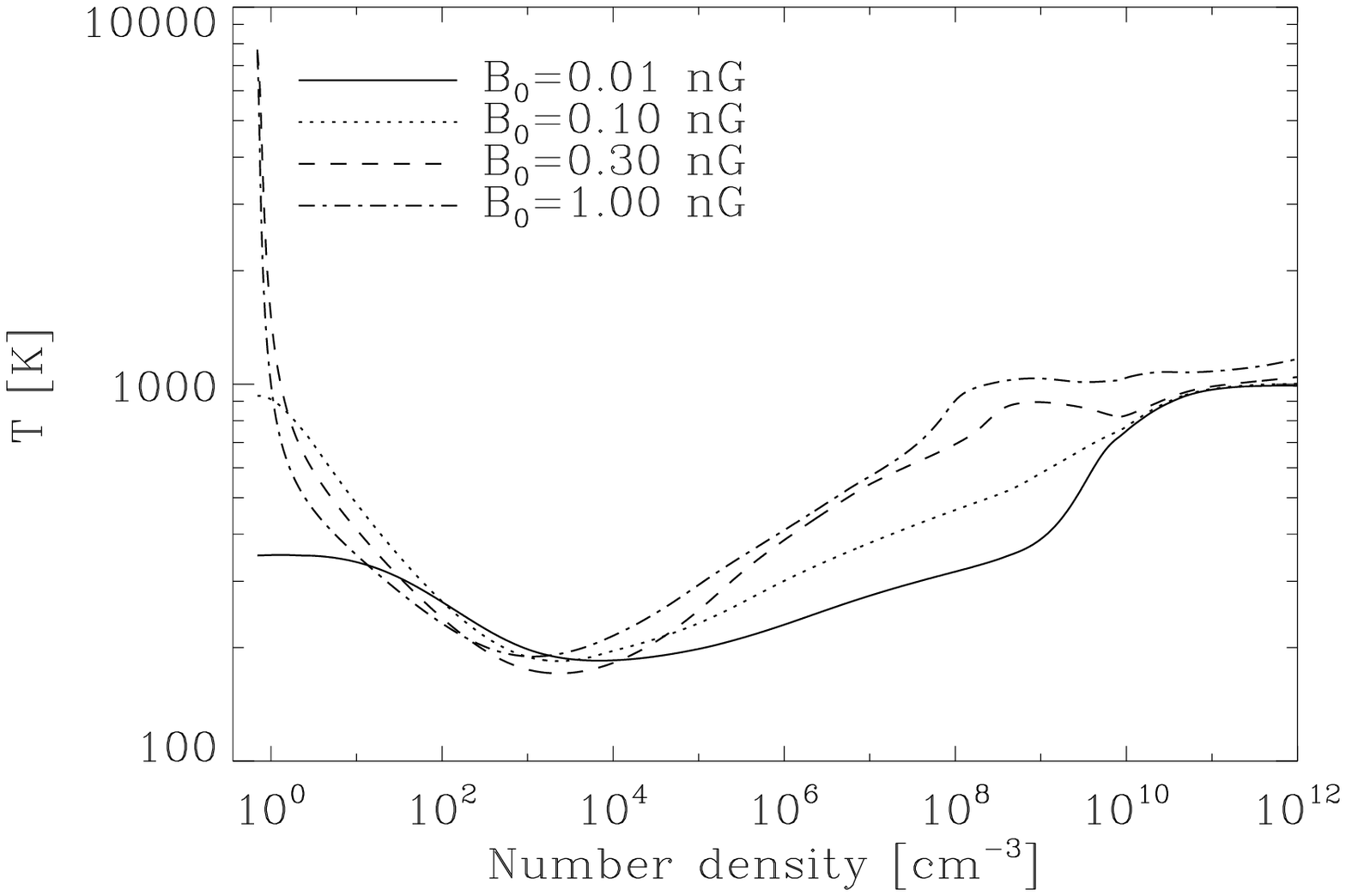}
\caption{The gas temperature as a function of density for different comoving field strengths. For $B_0=0.01$~nG, the thermal evolution corresponds to the zero-field case.}
\label{fig:tempcoll}
\eef


\begin{figure}[t]
\includegraphics[scale=0.5]{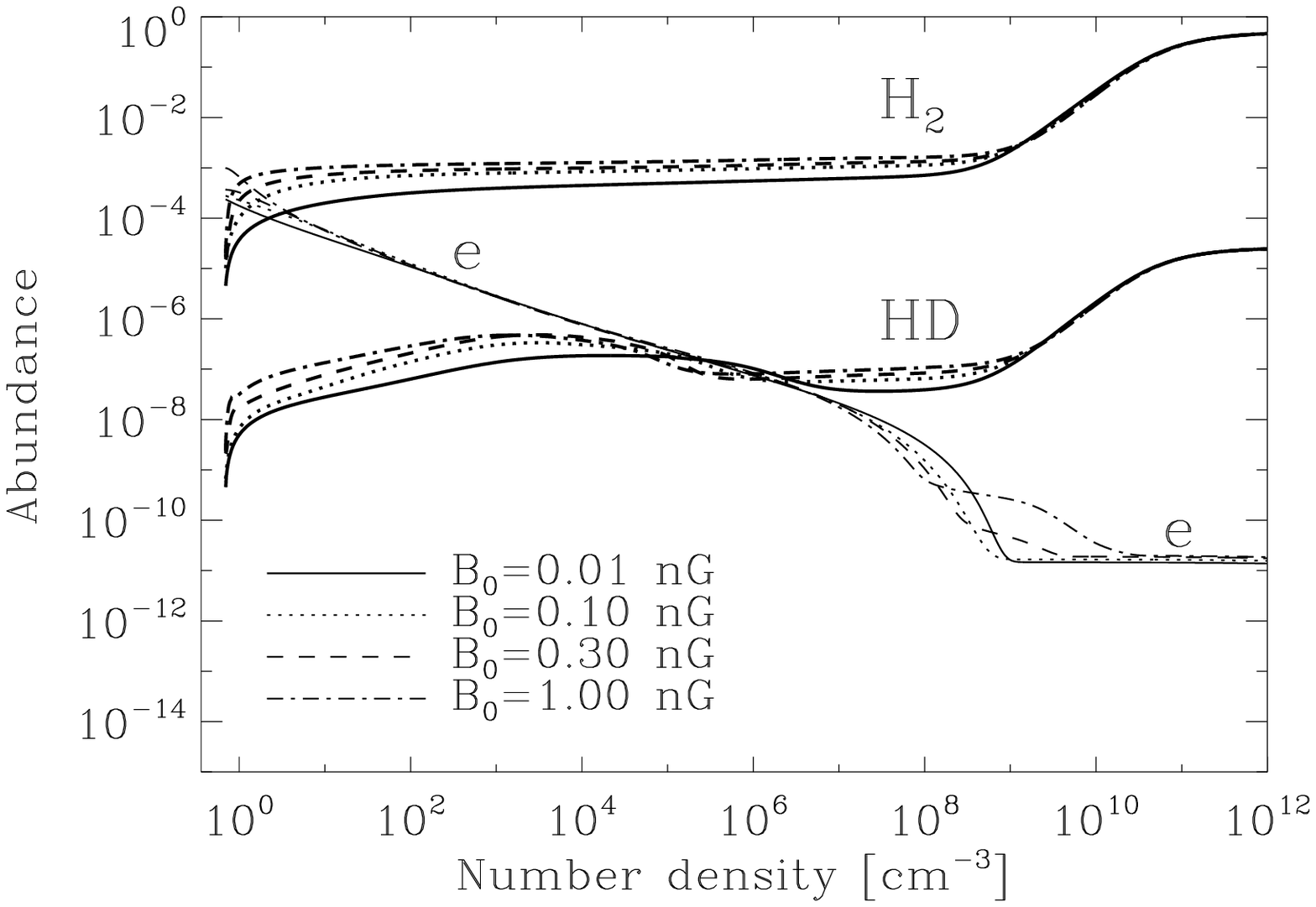}
\caption{Ionization degree, \HzI and \HDI abundance as a function of density for different comoving field strengths. For $B_0=0.01$~nG, the thermal evolution corresponds to the zero-field case. }
\label{fig:abuncoll}
\eef

\begin{figure}[t]
\includegraphics[scale=0.5]{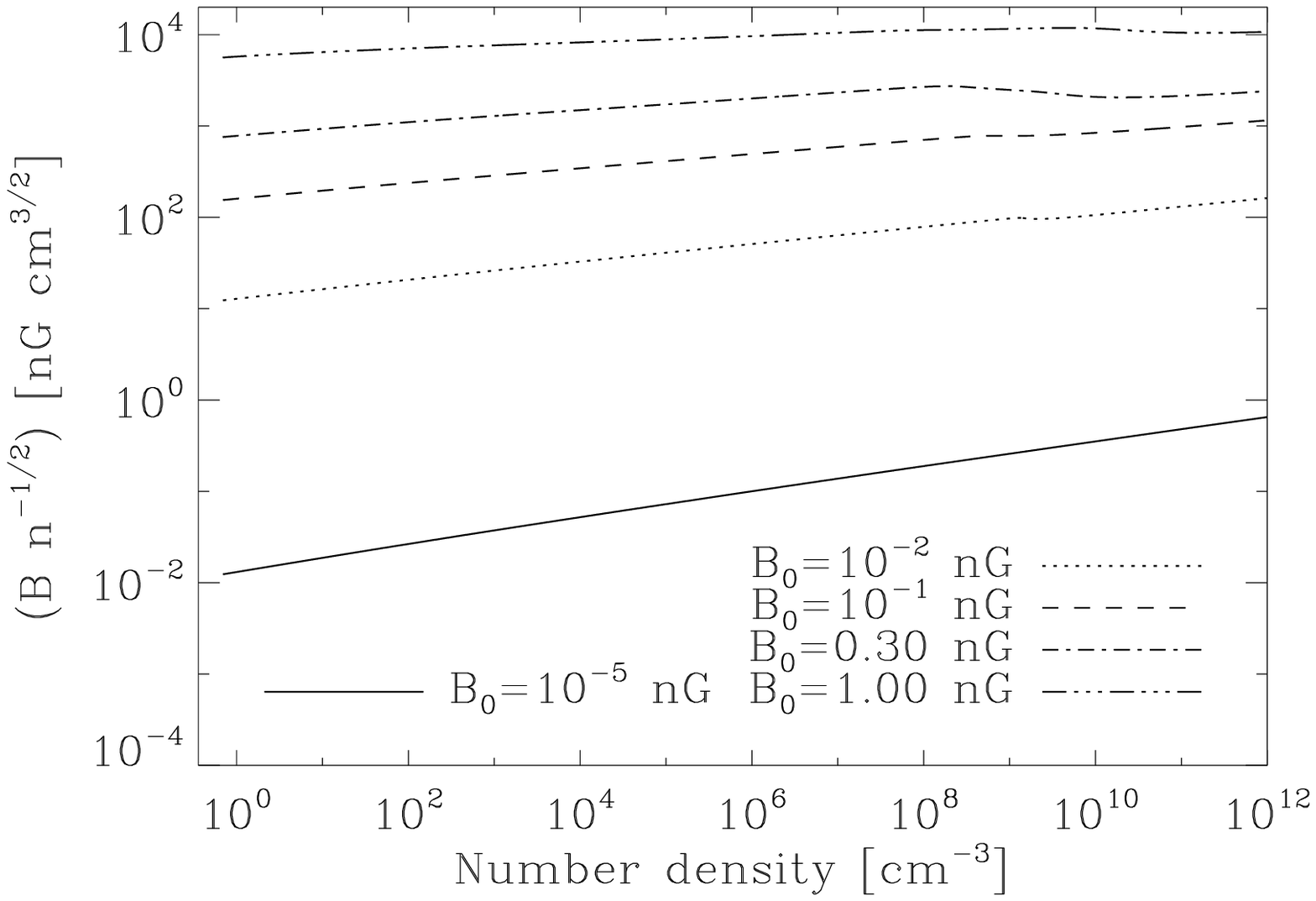}
\caption{The physical field strength as a function of density for different comoving field strengths.}
\label{fig:bfieldcoll}
\eef

\begin{figure}[t]
\includegraphics[scale=0.5]{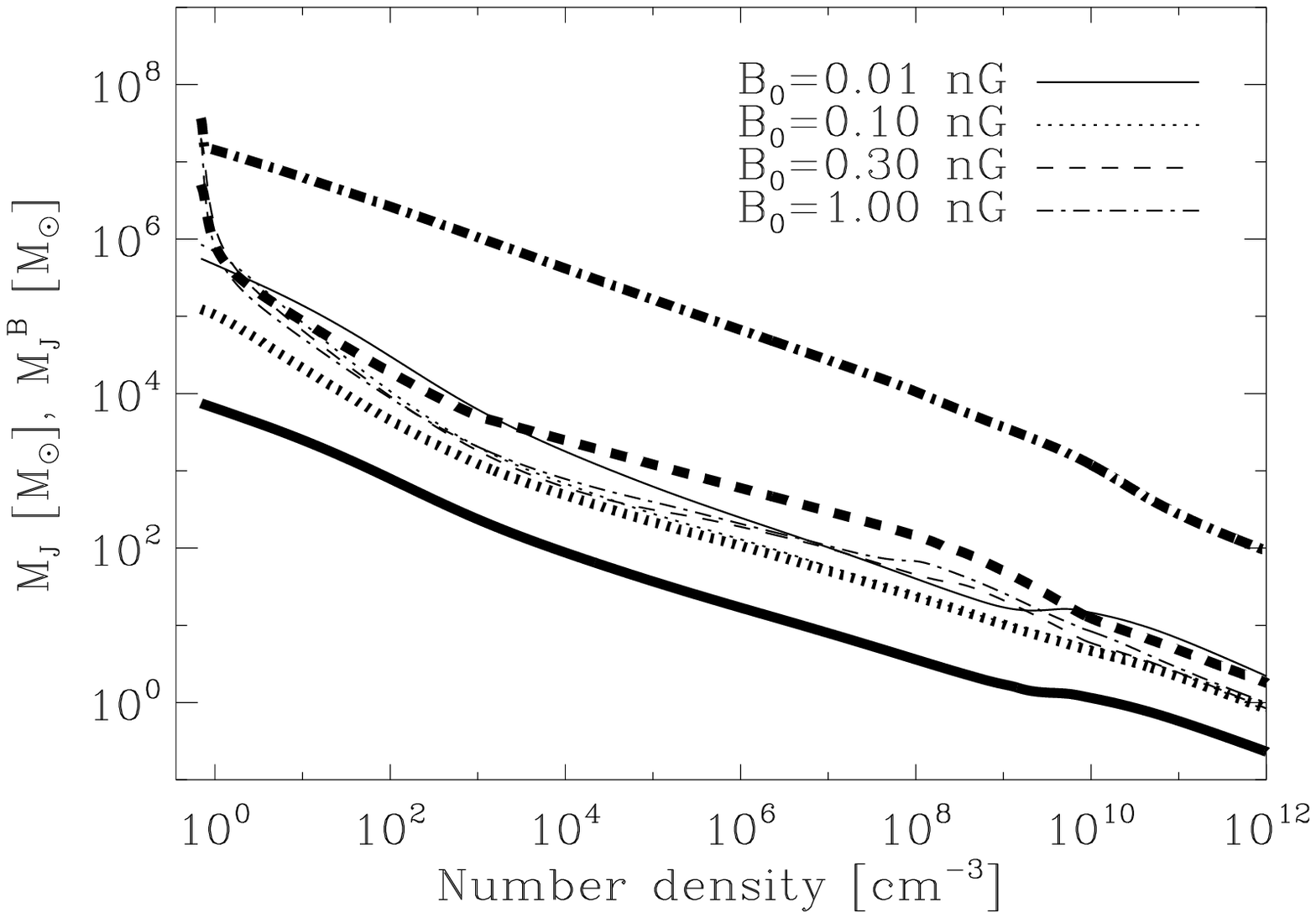}
\caption{Thermal (thin lines) and magnetic (thick lines) Jeans mass as a function of density for different comoving field strengths.  For $B_0=0.01$~nG, the thermal evolution and thus the thermal Jeans mass corresponds to the zero-field case.}
\label{fig:jeanscoll}
\eef

With the model described above, we calculate the evolution of the magnetic field strength and its impact on protostellar collapse in primordial gas. For this purpose, we adopt the results from \S~\ref{IGM} for the field strength, the temperature and the chemical abundances at virialization and use them as initial conditions for the collapse calculation. For comoving field strengths of at least $0.1$~nG, we showed in the previous section that the abundances of free electrons, \HzI and \HDI may be significantly increased in the IGM. During collapse, the enhanced fraction of free electrons may further catalyze the formation of molecules, so that the cooling rate increases in the presence of magnetic fields. On the other hand, there is also more heat input from magnetic energy dissipation. In the absence of magnetic fields, it was shown that such initial conditions may lead to cooling down to the CMB floor \citep{Yoshida07a, Yoshida07b} and characteristic stellar masses of $\sim10\ M_\odot$. We will assess here how these results change in the presence of magnetic energy dissipation. Finally, note that the label $B_0$ refers to the comoving field strength used to initialize the IGM calculation. The corresponding physical field strength at the beginning of the collapse phase is given in Table~\ref{tab:models}.

\begin{table}[htdp]
\begin{center}
\begin{tabular}{cc}
$B_0$~[nG] & $B$~[nG]  \\
\hline
$1$  &  $4.8\times10^3$ \\
$0.3$ & $6.3\times10^2$ \\
$0.1$ & $1.3\times10^2$ \\
$0.01$ & $1.0\times10^1$
\\
\hline
\end{tabular}
\end{center}
\caption{The physical field strength $B$ at beginning of collapse as a function of the comoving field strength $B_0$ used to initialize the IGM calculation at $z=1300$. 
}
\label{tab:models}
\end{table}%

Fig.~\ref{fig:tempcoll} shows the temperature evolution as a function of density for different comoving field strengths. For comoving fields of $0.01$~nG or less, there is virtually no difference in the temperature evolution from the zero-field case. For comoving fields of $\sim0.1$~nG, cooling wins over the additional heat input in the early phase of collapse, and the temperature decreases slightly below the zero-field value at densities of $10^3$~cm$^{-3}$. At higher densities, the additional heat input dominates over cooling and the temperature steadily increases. At densities of $\sim10^9$~cm$^{-3}$, the abundance of protons drops considerably and increases the AD resistivity defined in Eq.~(\ref{etaAD}) and the heating rate until Li$^+$ becomes the main charge carrier. In particular for comoving fields larger than $\sim0.1$~nG, this transition is reflected by a small bump in the temperature evolution due to the increased heating rate in this density range. The transition is also visible in Fig.~\ref{fig:bfieldcoll}, which shows the evolution of magnetic field strength with density. At the transition, we find a flattening of the relation between density and field strength, as a significant fraction of the energy can be dissipated at this stage. As expected from the evolution of the \HzI abundance, the main contribution to the total resistivity is due to the resistivity of atomic hydrogen, until the gas becomes fully molecular at densities of $\sim10^{11}$~cm$^{-3}$.

The evolution of the abundances of free electrons, \HzI and \HDI are given in Fig.~\ref{fig:abuncoll}. With higher initial field strength, the abundance of free electrons increases, inducing the formation of more molecules. The molecular abundances thus increase with field strength as well. Interestingly, the molecular fraction becomes independent of field strength at densities of $\sim10^{10}$~cm$^{-3}$ where three-body \HzI formation takes over. Note that there are still small differences in the electron abundance.

Apart from the transition where Li$^+$ becomes the dominant charge carrier, the magnetic field strength usually increases more rapidly than $\rho^{0.5}$, and weak fields increase more rapidly than strong fields. This is what one naively expects from Eq.~(\ref{fit}), and it is not significantly affected by magnetic energy dissipation. Another important point is that comoving fields of only $10^{-5}$~nG are amplified to values of $\sim1$~nG at a density of $10^3$~cm$^{-3}$. Such fields are required to drive protostellar outflows that can magnetize the IGM \citep{Machida06}. 

Fig.~\ref{fig:jeanscoll} shows the evolution of the thermal and magnetic Jeans mass during collapse. The thermal Jeans masses are quite different initially, but as the temperatures reach the same order of magnitude during collapse, the same holds for the thermal Jeans mass. The thermal Jeans mass in this late phase has only a weak dependence on the field strength. As expected, the magnetic Jeans masses are much more sensitive to the comoving field strength, and initially differ by about two orders of magnitude for one order of magnitude difference in the field strength. For comoving fields of $\sim1$~nG, the magnetic Jeans mass dominates over the thermal one and thus determines the mass scale of the protocloud. For $\sim0.3$~nG, both masses are roughly comparable, while for weaker fields the thermal Jeans mass dominates. The magnetic Jeans mass shows features both due to magnetic energy dissipation, but also due to a change in the thermal Jeans mass, which sets the typical length scale and thus the magnetic flux in the case that $M_J>M_J^B$.

\begin{figure}[t]
\includegraphics[scale=0.5]{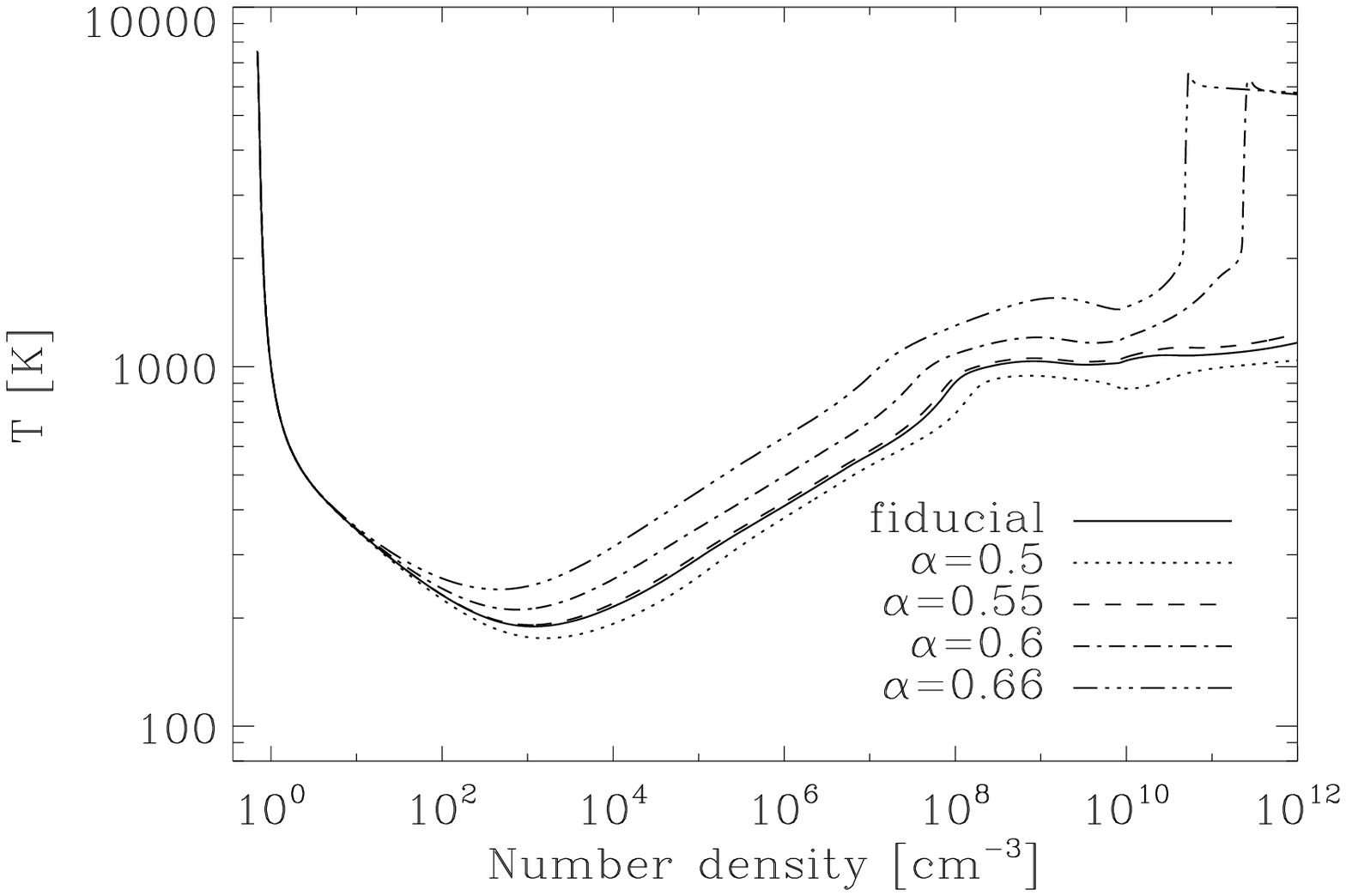}
\caption{The gas temperature as a function of density for a comoving field strength of $1$~nG and different assumptions for the scaling between density and field strength, both assuming constant slopes $\alpha$ and our fiducial model based on the fit given in Eq.~(\ref{fit}).}
\label{fig:tempcomp}
\eef

\begin{figure}[t]
\includegraphics[scale=0.5]{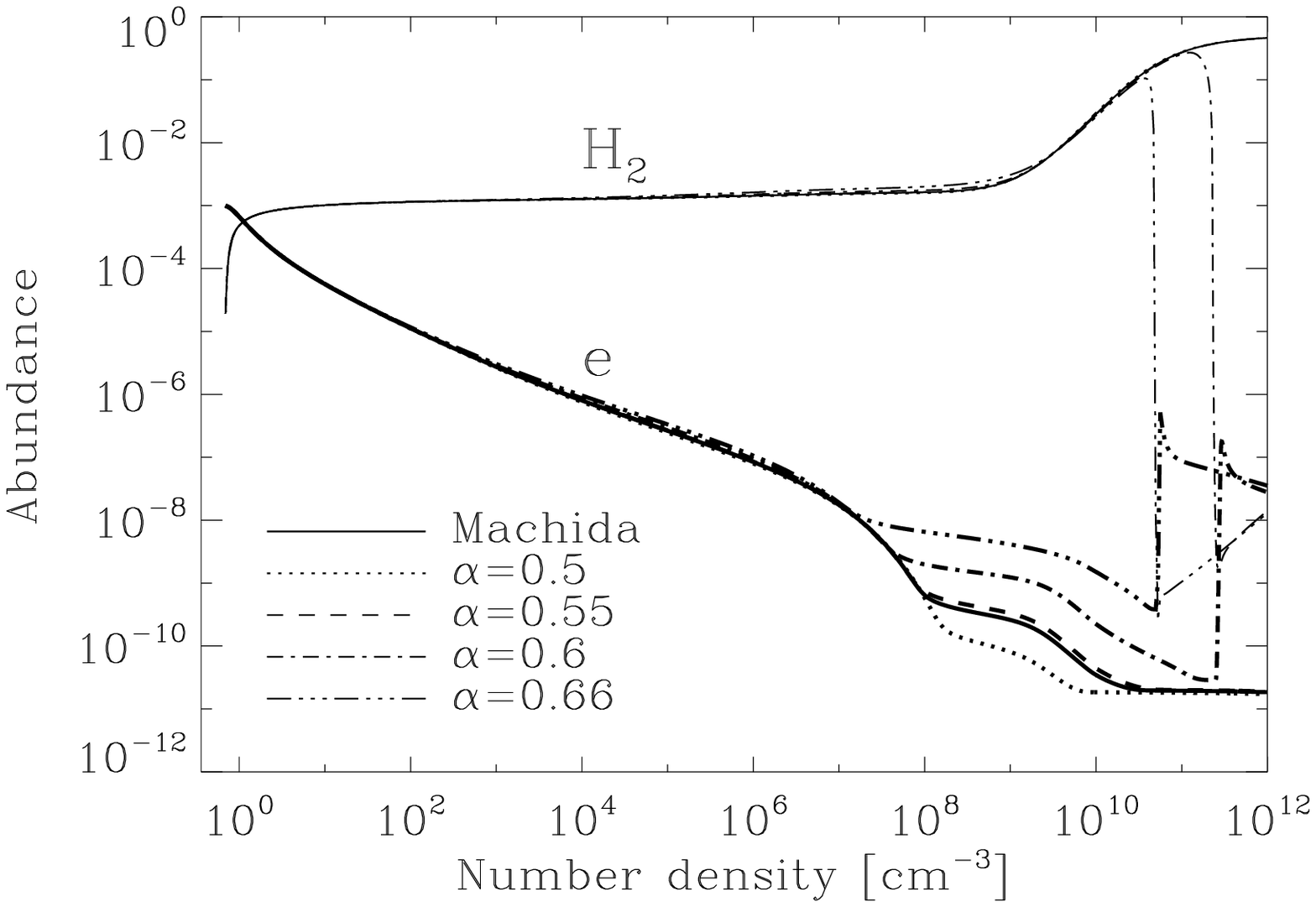}
\caption{Ionization degree (bold lines) and \HzI abundance (thin lines) as a function of density for a comoving field strength of $1$~nG and different assumptions for the scaling between density and field strength, both assuming constant slopes $\alpha$ and our fiducial model based on the fit given in Eq.~(\ref{fit}).}
\label{fig:abuncomp}
\eef

\begin{figure}[t]
\includegraphics[scale=0.5]{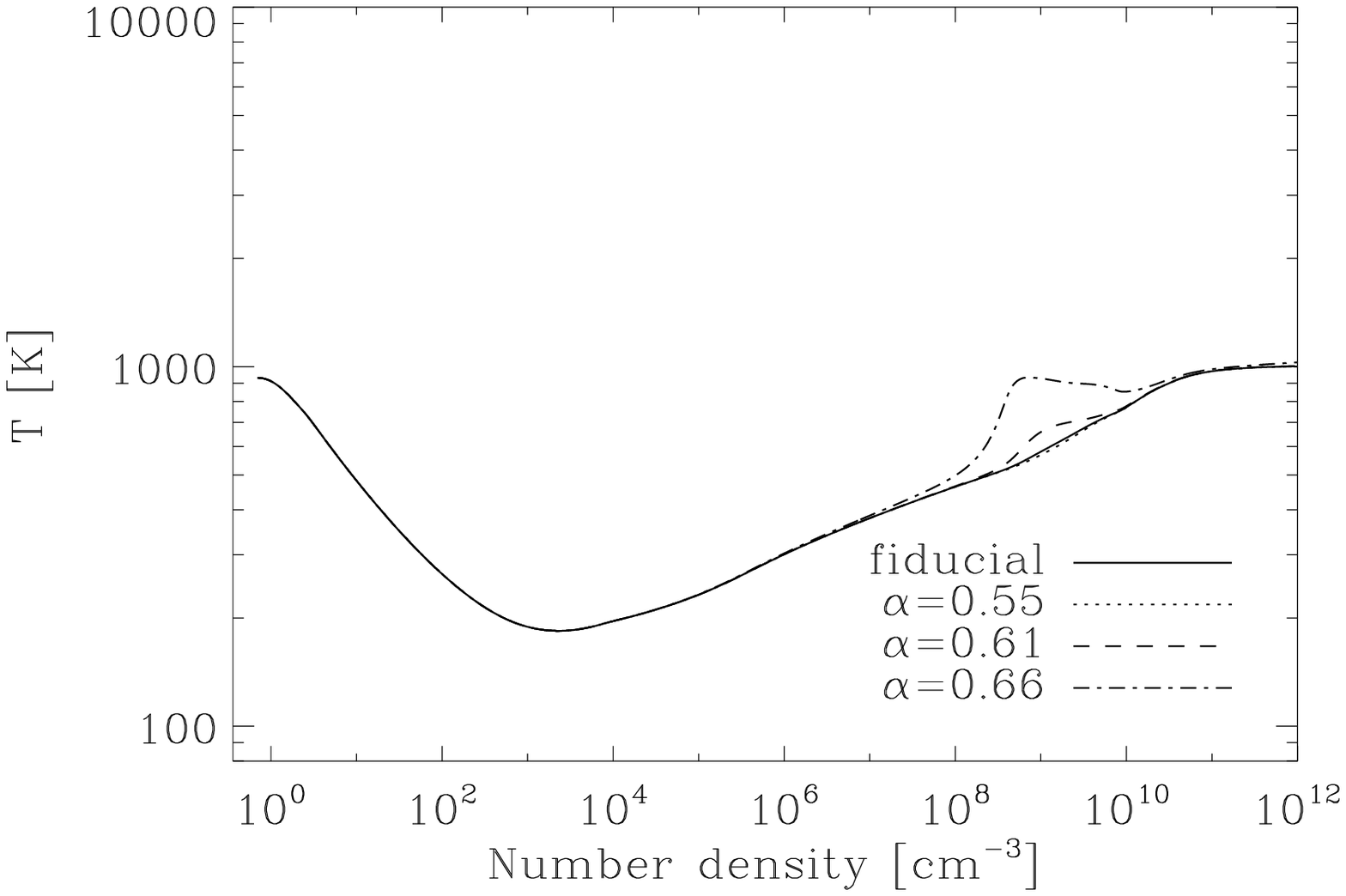}
\caption{The gas temperature as a function of density for a comoving field strength of $0.1$~nG and different assumptions for the scaling between density and field strength, both assuming constant slopes $\alpha$ and our fiducial model based on the fit given in Eq.~(\ref{fit}).}
\label{fig:tempcomp2}
\eef

We have checked the sensitivity of our results to our model assumptions, finding that the correct scaling relation between the magnetic field strength and density is crucial for these results. As an example, we show how the results for a $1$~nG field depend on this scaling relation in Figs.~(\ref{fig:tempcomp}) and (\ref{fig:abuncomp}). In this particular case, the fitting formula in Eq.~(\ref{fit}) matches best with a constant scaling $\alpha=0.55$. For larger values of $\alpha$, the heating rate increases rapidly with density, and reaches a critical value for which the \HzI dissociation rates become very effective. These dissociation rates change by several orders of magnitude in this temperature range are extremely sensitive to the temperature evolution. Once activated, they act to reduce the cooling from \HzId, which amplifies the temperature increase further and leads to the sudden jump shown in Fig.~(\ref{fig:tempcomp}). On the other hand, for $\alpha<0.55$, the temperature might be smaller than in our standard runs. It is interesting to note that the bump due to the transition where Li$^+$ becomes the main charge carrier becomes less prominent for steeper relations, as the latter considerably enhances the heating rate at higher densities.

 We have also checked the sensitivity to the scaling relation for a comoving field strength of $0.1$~nG (see Fig.~\ref{fig:tempcomp2}). In this case, the differences are smaller, but still significant at densities larger than $10^8$~cm$^{-3}$. For this initial field strength, the bump at $10^9$~cm$^{-3}$ becomes more prominent for a steeper relation. This is because AD heating is only significant for $\alpha>0.55$. For smaller values of $\alpha$, the differences with the zero-field case are mostly due to the change in the initial chemical abundances, and not due to the effects of AD during the collapse.

\section{Discussion}\label{discussion}
The calculations above show that comoving magnetic fields of $\sim0.1$~nG or more significantly increase the thermal and magnetic Jeans mass in the IGM to $10^7-10^9\ M_\odot$. The thermal evolution is modified both due to changes in the chemical initial conditions at the beginning of the collapse phase, as well as due to the further energy input from magnetic energy dissipation. A correct modeling of the AD resistivities is quite important, as these rise considerably at high densities when the proton abundance drops before Li$^+$ becomes the main charge carrier. Our results further indicate that comoving field strengths of $10^{-5}$~nG are sufficiently amplified during protostellar collapse to reach the critical field strength derived by \citet{Machida06} for the formation of jets, which may magnetize the IGM. {Of course, additional and perhaps even more important contributions may be generated once massive Pop.~III stars are formed \citep{Kogan73a,Kogan73b,Kogan77} .}

During protostellar collapse, we do not find a significant temperature change at the minimum near densities of $\sim10^3$~cm$^{-3}$, and therefore do not expect a large change in the characteristic fragmentation mass scale. At higher densities, the additional heat input will slow down the collapse, in particular at densities of $\sim10^9$~cm$^{-3}$ where Li$^+$ becomes the main charge carrier. This may lead to deviations from spherical symmetry and favor the formation of flattened structures like an accretion disk. Due to the higher temperatures, a higher accretion rate onto the protostar may be expected. Part of this may however be balanced by outflows induced by the magnetic fields, and the dynamical implications of our results need to be explored in three-dimensional MHD simulations.

{In summary, our results confirm that the presence of magnetic fields may increase the Jeans mass in the IGM considerably, which may suppress gravitational collapse in the first minihalos and therefore delay reionization. As shown by \citet{SchleicherBanerjee08}, it is therefore possible to derive upper limits on the comoving field strength, which are of the order of a few nG, depending on assumptions for the stellar population. With subsequent works, we plan to further reduce the uncertainties regarding the stellar population, such that tighter constraints can be derived. In addition, we expect that the Planck data and its more accurate measurement of the reionization optical depth will be very valuable in this respect. 
}

%


{It is also} of interest to understand how the effects of magnetic fields depend on the environmental conditions. For example, in relic HII regions and atomic cooling halos, the temperature and ionization fraction could be significantly increased, leading to a different evolution of the chemistry \citep{Yoshida07a, Yoshida07b}. We have checked that for comoving field strength of $\sim1$~nG, the additional heat input  from magnetic energy dissipation dominates over the additional cooling from the enhanced \HzI and \HDI abundance, while for comoving fields of $\lesssim0.1$~nG, the difference from the zero-field case is small, as the higher ionization degree makes AD less effective. Further effects may be due to the presence of turbulence in these massive atomic cooling halos \citep{Clark08, Greif08}. In particular, the coherence length of the magnetic field may be reduced increasing the AD heating rate, while energy dissipation from decaying MHD turbulence may deposit further heat into the gas.

Subsequent generations of stars may form from gas that has been enriched with metals, and presumably stronger magnetic fields due to stellar winds or supernova explosions. The impact of the additional metallicity has been discussed extensively in the literature \citep{Bromm01b, Schneider03, Omukai05, Schneider06, Glover07, Jappsen07, Jappsen07b, Jappsen09, Tornatore07, Clark08, Omukai08, Smith09}. Conversely, both the amount of magnetic energy generated in the IGM by the first generation of stars, as well as their possible effects on the subsequent generations are still largely unexplored \citep[but see][]{Rees87, Kandus04}.  {At low metallicity, dust cooling becomes important at high densities  \citep{Omukai05, Schneider06} and may easily dominate over the additional heating from ambipolar diffusion.  At lower densities, the main effect of dust is to increase the abundance of molecular hydrogen due to \HzI formation on dust grains \citep{Cazaux09}. This may also increase the cooling rate significantly. At the same time, the presence of metallic ions may increase the ionization fraction in the gas, such that ambipolar diffusion may become less important. However, it also implies that the dynamical effects of magnetic fields will play a more important role, as the coupling between the gas and the magnetic field will be more efficient.}

Finally, we stress a fundamental difference in the magnetically
controlled collapse of clouds of primordial and non-primordial chemical
compositions.  Whereas the electric charge at densities $n \gtrsim
10^9$~cm$^{-3}$ in a primordial cloud is mostly carried by Li$^+$ ions
and electrons (see Sect.~\ref{ResultsCollapse}), in a cloud of solar chemical composition
the charged component in the same density range is dominated by
negatively and positively charged dust grains \citep{Desch01, Nakano02}. This has important consequences on the evolution
of the magnetic field because the electric resistivities associated to
ambipolar diffusion, Hall effect, and Ohmic dissipation are expected to
differ even by orders of magnitude in the two cases, as shown e.g. by
\citet{Pinto08a}. Unfortunately, while the polarization
approximation adopted in our calculation of the AD resistivity
appears to be justified for collisions of Li$^+$ with He \citep{Cassidy85} and H$_2$ \citep{Roeggen02, Dickinson82}, collisions of Li$^+$ with H atoms, especially relevant for
the primordial gas, have not been studied in detail either theoretically or
experimentally.

As noted above, all of these results are sensitive to the correct scaling relation between the magnetic field strength and the gas density. A correct and self-consistent derivation of this relation is therefore important for our understanding of primordial star formation.

\acknowledgments

D.R.G.S. thanks the Heidelberg Graduate School of Fundamental Physics (HGSFP) and the
State of Baden-W{\"u}rttemberg for financial support. R.S.K. acknowledges support from
the German Science Foundation (DFG) under the Emmy Noether grant KL1358/1 and the
Priority Program SFB 439 Galaxies in the Early Universe. S.C.O.G. acknowledges funding
from the DFG via grant KL1358/4. RB is funded by the DFG under the grant BA 3706/1. This work was supported in part by a FRONTIER grant of
Heidelberg University sponsored by the German Excellence Initiative. We also acknowledge funding from the European Commission Sixth Framework Programme Marie Curie Research Training Network CONSTELLATION (MRTN-CT-2006-035890). {We thank the anonymous referee for many clarifying remarks that helped to improve this paper.}


\clearpage




\clearpage

\end{document}